\documentclass[showpacs,prd,amsfonts,amsmath,twocolumn,aps,preprintnumbers,letterpaper]{revtex4}
\usepackage{amsmath,amssymb}
\usepackage{epsfig}
\usepackage{graphicx}
\usepackage{tabularx}
\usepackage{amsmath}
\usepackage{slashed}
\usepackage{amsfonts}
\usepackage{epstopdf}
\usepackage{color}
\usepackage[caption=false]{subfig}
\usepackage[pdftex, pdfborder= 0 0 0, citecolor=blue, urlcolor=blue, linkcolor=blue, colorlinks=true, bookmarksopen=true]{hyperref}

\newcommand{\be}{\begin{equation}}
\newcommand{\ee}{\end{equation}}
\newcommand{\bear}{\begin{eqnarray}}
\newcommand{\eear}{\end{eqnarray}}
\newcommand{\ba}{\begin{array}}
\newcommand{\ea}{\end{array}}

\def\be{\begin{eqnarray}}
\def\ee{\end{eqnarray}}
\def\bea{\be}
\def\eea{\ee}

\def\roughly#1{\mathrel{\raise.3ex\hbox{$#1$\kern-.75em%
\lower1ex\hbox{$\sim$}}}}

  \long\def\comment#1{ }

  \newcommand{\beq}{\begin{eqnarray}}
  \newcommand{\eeq}{\end{eqnarray}}
  
 \def\simge{\mathrel{%
   \rlap{\raise 0.511ex \hbox{$>$}}{\lower 0.511ex \hbox{$\sim$}}}}
\def\simle{\mathrel{
   \rlap{\raise 0.511ex \hbox{$<$}}{\lower 0.511ex \hbox{$\sim$}}}}

\begin{document}

\title{Electromagnetic radii of the nucleon  in soft-wall holographic QCD}

\author{Kiminad A. Mamo}
\email{kmamo@anl.gov}
\affiliation{Physics Division, Argonne National Laboratory, Argonne, Illinois 60439, USA
}

\author{Ismail Zahed}
\email{ismail.zahed@stonybrook.edu}
\affiliation{Center for Nuclear Theory, Department of Physics and Astronomy, Stony Brook University, Stony Brook, New York 11794-3800, USA}

\date{\today}

\begin{abstract}
We revisit the electromagnetic form factors of the proton and neutron in the original-minimal soft-wall holographic QCD,  which  has only two parameters, i.e., the mass scale $\kappa$ and the twist parameter of the nucleon $\tau$. We first fix $\tau=3$ by the hard scattering rule, and extract $\kappa=0.402~\text{GeV}$ from the world data (including the Mainz A1 data) of the Sachs magnetic form factor of the proton $G_M^p$. We then predict among others, the charge radius of the proton to be $\bf{r_{p}=0.831\,\pm\,0.008\,\,\text{fm}}$, in perfect agreement with the recent charge radius of the proton measured by the PRad collaboration at Jefferson Lab, and in agreement with the muonic hydrogen experiments. Our prediction for the proton elastic form factor ratio $\mu_pG_E^p/G_M^p$ is also in very good agreement with the recent high precision Jefferson Lab recoil polarization experiment E08-007 for $Q^2=0.3\,-\,0.6~\text{GeV}^2$, and with the recent high precision Mainz A1 experiment for $Q^2< 0.13~\text{GeV}^2$.  
\end{abstract}

\maketitle

%

\section{Introduction}
\label{Introduction}

The nucleon is a composite hadron with quarks and gluons constituents, the stuff at the origin of all hadronic matter.
The quarks are charged and their distribution inside a nucleon is captured by the electric and magnetic charge radii,
which  measure the charge and current distributions respectively.  Precision electron scattering and spectroscopic 
measurements show a stubborn  4\% discrepancy, the so-called proton radius puzzle~\cite{Carlson:2015jba,Bernauer:2020ont,Gao:2021sml,Arrington:2021alx}. 
This is a surprising state of affairs given the fundamental nature of the proton. 

\begin{figure}[!htb]
\includegraphics[height=4.2cm,width=7.8cm]{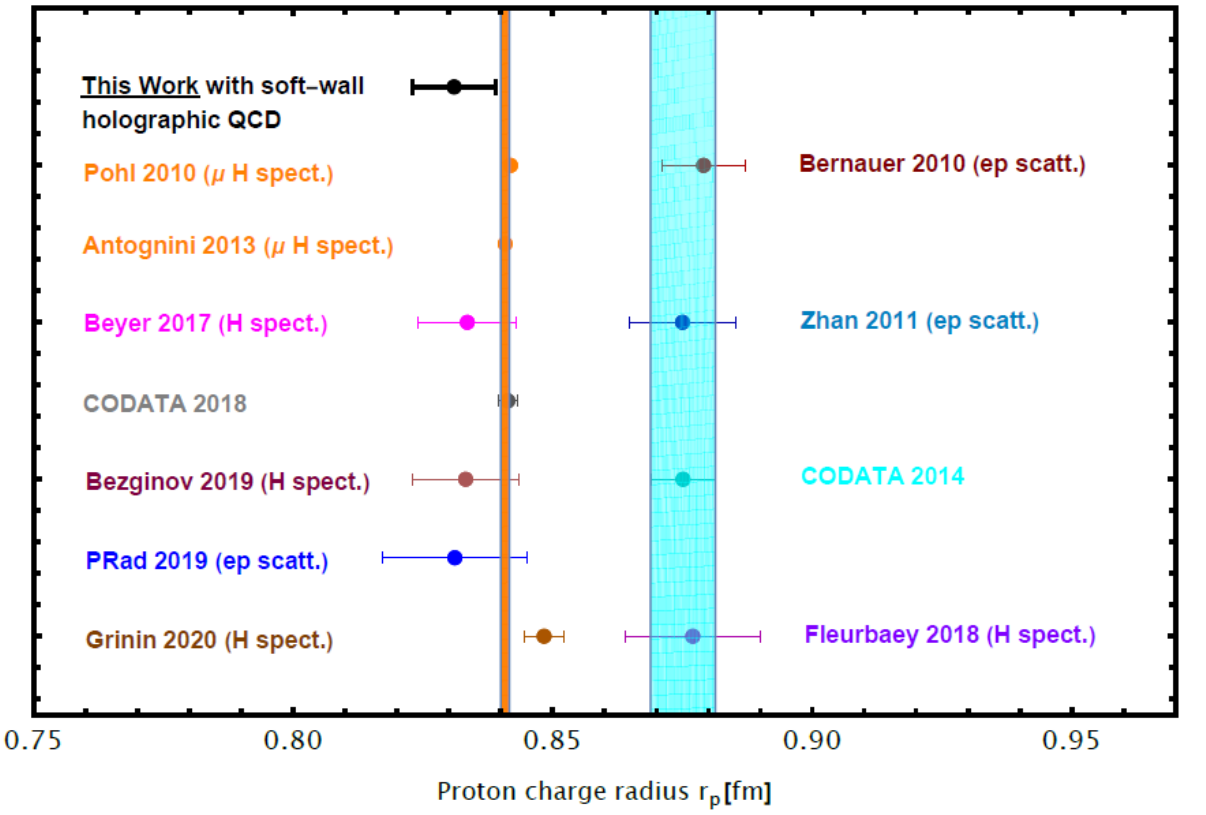}
 \caption{World data for the proton  charge radius. The black dot (with its theoretical error bar) is our prediction within the original-minimal soft-wall holographic QCD \cite{Karch:2006pv} with $\tau=3$ and $\kappa=0.402~\rm GeV$, see Table~\ref{table11}. The dark-yellow dot is the prediction of Abidin and Carlson \cite{Abidin:2009hr} (which is also within the original-minimal soft-wall holographic QCD \cite{Karch:2006pv} with $\tau=3$ and $\kappa=0.350~\rm GeV$), see also Table~\ref{table11}. The other dots are the world experimental data for the proton charge radius Pohl 2010 ($\mu$ H spect.) \cite{Pohl:2010zza}, Antognini 2013 ($\mu$ H spect.) \cite{Antognini:2013txn}, Beyer 2017 (H spect.) \cite{CREMA_2017}, CODATA 2018 \cite{CODATA_2018}, Bezignov 2019 (H spect.) \cite{Bezginov:2019mdi}, PRad 2019 (ep scatt.) \cite{Xiong:2019umf}, Grinin 2020 (H spect.) \cite{Grinin20}, Bernauer 2010 (ep scatt.) \cite{Bernauer:2010wm}, Zhan 2011 (ep scatt.) \cite{Zhan:2011ji}, CODATA 2014 \cite{CODATA_2014}, and Fleurbaey 2018 (H spect.) \cite{Fleurbaey:2018fih}.
}
  \label{fig_data11}
\end{figure}

The proton structure is the quintessential QCD problem, currently addressed using  
 ab-initio lattice simulations~\cite{Durr:2008zz}.  Empirically, the mass of the proton is known with great accuracy, but its fundamental
charge radius is not, as shown in Fig.~\ref{fig_data11}.  The high precision elastic e-p scattering measurement of 0.879 fm by Mainz A1 collaboration~\cite{Bernauer:2010wm} while
consistent with the value of 0.877 fm from the previous hydrogen spectroscopy~\cite{Fleurbaey:2018fih}, is larger than the reported value of 0.841 fm from muonic hydrogen spectroscopy \cite{Antognini:2013txn}, and the values of 0.833 fm and 0.848 fm from the recent hydrogen spectroscopy measurements~\cite{Bezginov:2019mdi,Grinin20}. Moreover, the newest e-p measurement by the PRad collaboration at Jefferson Lab has reported a small charge radius of 0.831 fm~\cite{Xiong:2019umf}. A smaller charge radius is also supported by various re-analysis of the Mainz A1 e-p scattering data~\cite{Mart:2013gfa,Lorenz:2014vha,Griffioen:2015hta,Higinbotham:2015rja,Horbatsch:2016ilr,Zhou:2018bon} with few exceptions \cite{Lee:2015jqa,Gramolin:2021gln}. In addition, the recent combined re-analysis of the Mainz A1 and PRad data has also favored a small charge radius \cite{Alarcon:2020kcz,Atac:2020hdq,Lin:2021umk,Cui:2021vgm,Zhou:2021gyh}.

All in all,  tension seems to exist between e-p scattering measurements of the proton charge radius, and atomic
measurements using muonic hydrogen as illustrated in Fig.~\ref{fig_data11}. Although ideas beyond the standard model have been suggested 
to fix the discrepancy~\cite{Krasznahorkay:2015iga,Carlson:2012pc}, 
none so far has been empirically conclusive. 

The proton structure emerges  from a subtle interplay between the
 sources composing the nucleon and the primordial glue in the vacuum~\cite{Zahed:2021fxk}.
 Most of the proton mass arises from the primordial and topological glue at the origin of the spontaneous breaking of chiral symmetry, but
its moderatly large size makes it still susceptible to the effects of confinement. This last 
feature is also shared by the  vector mesons, but not the light and more compact pseudoscalar Goldstone
bosons. This last observation is further supported by the straight character of the nucleon 
and rho Regge trajectories,  and their remarkable parallelism.

We will address the proton and neutron  size problem non-perturbatively in the context of holographic QCD, 
which among other embodies Regge physics,  and provides a field theoretical realization  for the dual 
resonance framework~\cite{Frampton:1969ry} postulated decades ago outside the realm of QCD.
It is the most economical way of enforcing basic symmetries, including crossing symmetry and unitarity,
the key tenets in dispersive analyses. 
The approach, originates from a conjecture that observables in 
strongly coupled and conformal gauge theories in the limit of a large number of colors and strong gauge coupling,
 can be determined from classical  fields interacting through gravity,  in an anti-de-Sitter space in 
 higher dimensions~\cite{Maldacena:1997re,Aharony:1999ti}. The conjecture has been extended since to non-conformal gauge theories~\cite{Witten:1998zw,Karch:2006pv}.

\begin{figure*}
\subfloat[\label{fig_3pt1}]{%
  \includegraphics[height=5.5cm,width=.46\linewidth]{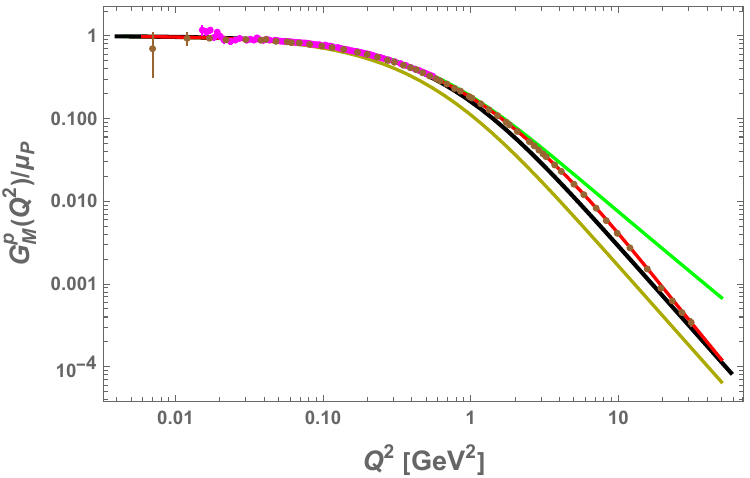}%
}\hfill
\subfloat[\label{fig_3pt6}]{%
  \includegraphics[height=5.5cm,width=.46\linewidth]{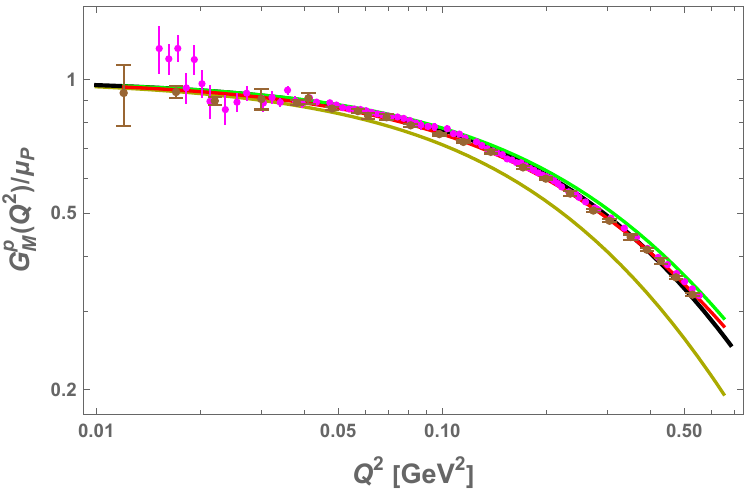}%
}\hfill
\subfloat[\label{fig_4pt1}]{%
  \includegraphics[height=5.5cm,width=.46\linewidth]{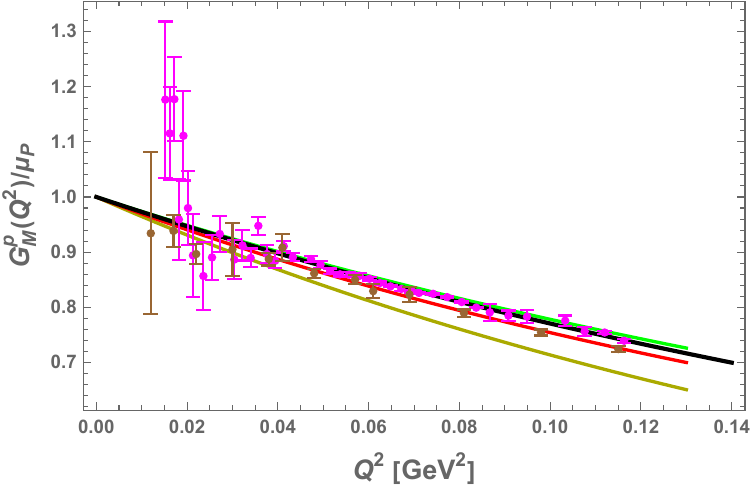}%
}
\caption{Sachs magnetic form factor of the proton at high and low momentum transfer. 
The black-solid curve is our soft-wall holographic QCD result with $\tau=3$ (fixed by the hard counting rule) and $\kappa=0.402\,\text{GeV}$ (extracted from the best fit to the world data (including Mainz A1 data) of Sachs magnetic form factor of proton). The green-solid curve is our soft-wall holographic QCD result with $\tau={m_N^2}/{m_{\rho}^2}+1=2.465$ and $4\kappa^2=m_{\rho}^2=0.775^2\,\text{GeV}^2$.  
The  dark-yellow-solid curve is the soft-wall holographic QCD result for $\tau=3$ and $\kappa=0.350~\rm GeV$~\cite{Abidin:2009hr}. The solid-red curve is the Arrington fit to world data (without Mainz A1 data)~\cite{Arrington:2007ux} (brown data points). The magenta data points are the Mainz A1 data~\cite{Bernauer:2013tpr}. Throughout, we have used the proton magnetic moment $\mu_p=G_{M}^{P}(0)=F_{1}^{P}(0)+F_{2}^{P}(0)=2.793$.}
\label{fig_GMP}
\end{figure*}

The holographic description of the electromagnetic form factors in the context of the original-minimal soft-wall holographic QCD \cite{Karch:2006pv} was initially addressed in~\cite{Abidin:2009hr} (with just two parameters, the mass scale $\kappa$ and twist of proton $\tau$) with rather large electromagnetic radii for $\kappa=0.350~\text{GeV}$ (fixed by simultaneously fitting the holographic mass of the 
rho meson and proton to their experimental values, see Table~\ref{table11}), and $\tau=3$ (fixed by the hard scattering rule). Here we show how
to overcome this major shortcoming by fixing $\kappa=0.402~\text{GeV}$ (fixed by fitting the holographic Sachs magnetic form factor of the proton to the world data (including the Mainz A1 data) of the Sachs magnetic form factor of proton), see Table~\ref{table11}), and Fig.~\ref{fig_GMP}. We also fix $\kappa=m_{\rho}/2=0.388~\text{GeV}$, and $\tau={m_N^2}/{m_{\rho}^2}+1=2.465$ at low energy, assuming that the nucleon anomalous dimension or  the twist $\tau$,  runs with the energy scale to asymptote its hard scaling value $\tau=3$ at high energy~\cite{Lepage:1980fj}, see Table~\ref{table11}. The difference between these two choices of the holographic parameters $\kappa$ and $\tau$,  will help us estimate our theoretical uncertainty in determining the electromagnetic radii of the nucleons.

The organization of the paper is as follows: in section~\ref{EMSachs}, we write down the holographic Sachs electric and magnetic form factor of the nucleon which are derived in detail, within the original-minimal holographic QCD \cite{Karch:2006pv,Abidin:2009hr}. In section~\ref{kappatau}, we show the three possible ways of fixing the only two parameters $\kappa$ and $\tau$ of the original-minimal holographic QCD, and summarize the corresponding predictions for the charge radius of the proton in Table~\ref{table11}. Finally, in section~\ref{EMRadii}, we compute the charge and magnetic radii of both the proton and neutron, and compare the holographic predictions, within the theoretical uncertainties, to the experimental values. We also summarize our results in Table~\ref{table22}. Our conclusions are in~\ref{Conclusion}.  More details can be found in the Appendices.


\section{Nucleon electric and magnetic Sachs form factors}\label{EMSachs}
We define the standard electric and magnetic Sachs form factors of the proton and neutron as
\be
G_E^{P,N}(Q)&=&F_1^{P,N}(Q)-\frac{Q^2}{4m_N^2}F_2^{P,N}(Q)\,,\label{electric}\\
G_M^{P,N}(Q)&=&F_1^{P,N}(Q)+F_2^{P,N}(Q)\,,\label{magnetic}
\ee
where the Dirac $F_1(Q)$ and Pauli $F_2(Q)$ form factors of proton (P) and neutron (N) in the original-minimal soft-wall holographic QCD \cite{Karch:2006pv} are given by (see the Appendix for their detailed derivation, see also \cite{Abidin:2009hr,Vega:2010ns})
\begin{widetext}
\bea
\label{F1P}
F_{1}^{P}(Q)&=&\left(\frac{a}{2}+\tau \right)\times B(\tau ,a+1)+\eta_{P}\times (\tau -1)\times \frac{a (a (\tau -1)-1)}{(a+\tau ) (a+\tau +1)}\times B(\tau -1,a+1)\,,\\
\label{F2P}
F_{2}^{P}(Q)&=&\eta_{P}\times 4(\tau -1)\times \tau \times B(\tau ,a+1)\,,\\
\label{F1N}
F_{1}^{N}(Q)&=&\eta_{N}\times (\tau -1)\times \frac{a (a (\tau -1)-1)}{(a+\tau +1) (a+\tau )}\times B(\tau -1,a+1)\,,\\
\label{F2N}
F_{2}^{N}(Q)&=&\eta_{N}\times 4(\tau -1)\times \tau \times B(\tau ,a+1)\,,
\eea
Here we have defined  $a=Q^2/4\kappa^2$ with $Q^2=-q^2\leq 0$,  and denoted by $\kappa$  the mass scale, $\tau$ the twist parameter of the nucleon, and by 
$B(x,y)={\Gamma(x)\Gamma(y)}/{\Gamma(x+y)}$  the Euler beta function. The additional coefficients  $\eta_{P,N}=2g_5^2\times(\eta_3 \pm  \eta_0)/2$ in (\ref{F1P}-\ref{F2N}) follow from the extra couplings $\eta_{0,3}$, for the singlet and triplet contributions of the bulk Pauli contribution~\cite{Abidin:2009hr}. 

The Pauli parameter of the proton $\eta_P$ can be determined by matching the value of $F_2^{P}(0)$ with the experimental data which is $1.793$, i.e.
\be
\eta_P=\frac{1.793}{C_3(0)}=\frac{1.793}{4 (\tau -1)}\,.
\ee
Similarly, the Pauli parameter of the neutron $\eta_N$ can be determined by matching the value of $F_2^{N}(0)$ with the experimental data which is $-1.913$, i.e.
\be
\eta_N=\frac{-1.913}{C_3(0)}=\frac{-1.913}{4 (\tau -1)}\,.
\ee

\begin{table}[h!]
\setlength{\arrayrulewidth}{0.1mm}
\setlength{\tabcolsep}{18pt}
\renewcommand{\arraystretch}{1.5}
\centering
\begin{tabular}{
 || m{3cm} 
 || m{1cm}
  | m{0.25cm}
  | m{8em}
  | m{8em}
  |  m{1.5cm} || }
\hline\hline
Soft-wall AdS/QCD & $\kappa$ & $\tau$ & $m^2_{\rho}=4\kappa^2$ & $m^2_{N}=4\kappa^2(\tau-1)$  & $\sqrt{\langle r^p_C\rangle^2}$ \\
\hline\hline
This Work &  0.388\,GeV & 2.465 & $0.775^2$\,$\text{GeV}^2$\,\,\, (\textbf{exp.}) & $0.938^2$\,$\text{GeV}^2$\,\,\, (\textbf{exp.}) &  0.839\,\,\,fm (\textbf{pr.})   \\
\hline
This Work &  0.402\,GeV & 3.000 & $0.804^2$\,$\text{GeV}^2$\,\,\,  (\textbf{pr.})   & $1.137^2$\,$\text{GeV}^2$\,\,\, (\textbf{pr.}) &  0.831\,\,\,fm (\textbf{pr.})   \\
\hline
Abidin-Carlson \cite{Abidin:2009hr}  &  0.350\,GeV & 3.000 & $0.700^2$\,$\text{GeV}^2$\,\,\, (\textbf{pr.}) & $0.990^2$\,$\text{GeV}^2$\,\,\, (\textbf{pr.}) &  0.960\,\,\,fm (\textbf{pr.})   \\ 
\hline\hline
\end{tabular}
\caption{Comparison of the values of the holographic parameters ($\kappa$ and $\tau$), and the corresponding predictions for the charge radius of the proton $\sqrt{\langle r^p_C\rangle^2}$ in our work (see (\ref{crp22}\,-\,\ref{crp33})) within the original-minimal soft-wall holographic QCD \cite{Karch:2006pv}, and the work of Abidin and Carlson \cite{Abidin:2009hr} (which is also within the original-minimal soft-wall holographic QCD \cite{Karch:2006pv}). Note that \textbf{pr.} and \textbf{exp.} are shorthand for  the \textbf{predicted} and \textbf{experimental} values, respectively.}\label{table11}
\end{table}
\end{widetext}

\begin{figure*}
\subfloat[\label{fig_3pt1}]{%
  \includegraphics[height=5.5cm,width=.46\linewidth]{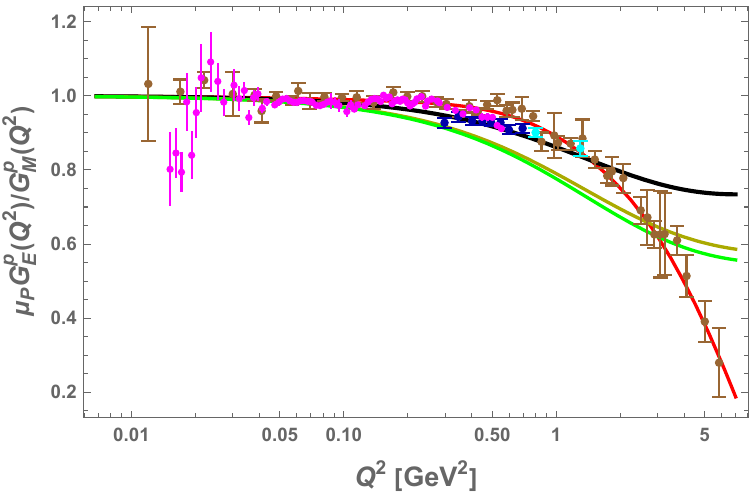}%
}\hfill
\subfloat[\label{fig_3pt6}]{%
  \includegraphics[height=5.5cm,width=.46\linewidth]{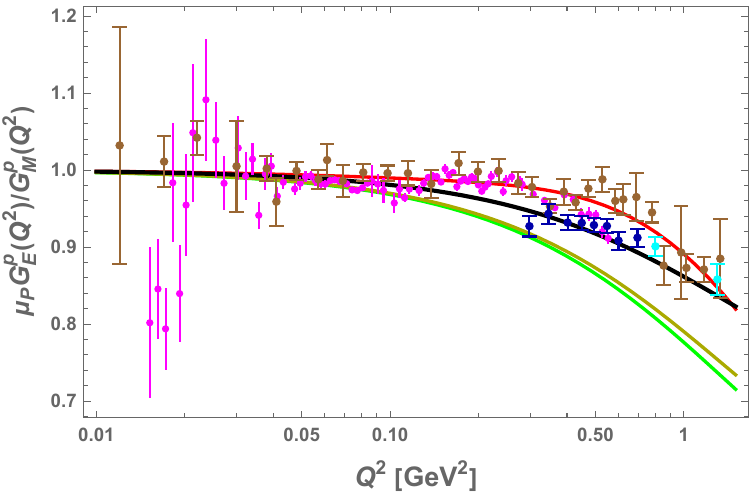}%
}\hfill
\subfloat[\label{fig_4pt1}]{%
  \includegraphics[height=5.5cm,width=.46\linewidth]{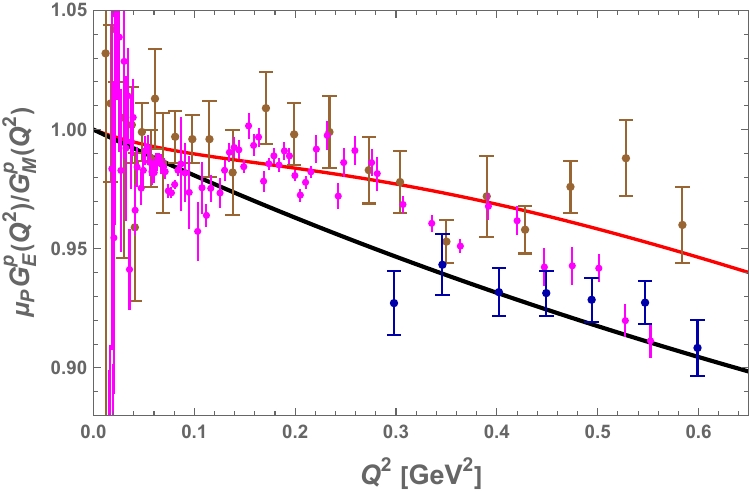}%
}
\caption{The ratio of the Sachs electric and magnetic form factor of the proton at high and low momentum transfer. 
The black-solid curve is our soft-wall holographic QCD result with $\tau=3$ (fixed by the hard counting rule) and $\kappa=0.402\,\text{GeV}$ (extracted from the best fit to the world data (including Mainz A1 data) of Sachs magnetic form factor of proton). The green-solid curve is our soft-wall holographic QCD result with $\tau={m_N^2}/{m_{\rho}^2}+1=2.465$ and $4\kappa^2=m_{\rho}^2=0.775^2\,\text{GeV}^2$.  
The  dark-yellow-solid curve is the soft-wall holographic QCD result for $\tau=3$ and $\kappa=0.350~\rm GeV$~\cite{Abidin:2009hr}. The solid-red curve is the Arrington fit \cite{Arrington:2007ux} to the world data~(brown data points) (note that the world data (brown data points) is without the Mainz A1 \cite{Bernauer:2013tpr} and the recent JLab recoil polarization experimental data \cite{Paolone:2010qc,Zhan:2011ji}). The magenta data points are the Mainz A1 data~\cite{Bernauer:2013tpr}.  The two cyan data points are from the recent high precision JLab recoil polarization experiment E03-104~\cite{Paolone:2010qc}. The dark-blue data points are from the recent high precision JLab recoil polarization experiment E08-007~\cite{Zhan:2011ji}. Throughout, we have used the proton magnetic moment $\mu_p=G_{M}^{P}(0)=F_{1}^{P}(0)+F_{2}^{P}(0)=2.793$.}
\label{fig_GEPGMP}
\end{figure*}

\begin{figure*}
\subfloat[\label{fig_3pt1}]{%
  \includegraphics[height=5.5cm,width=.46\linewidth]{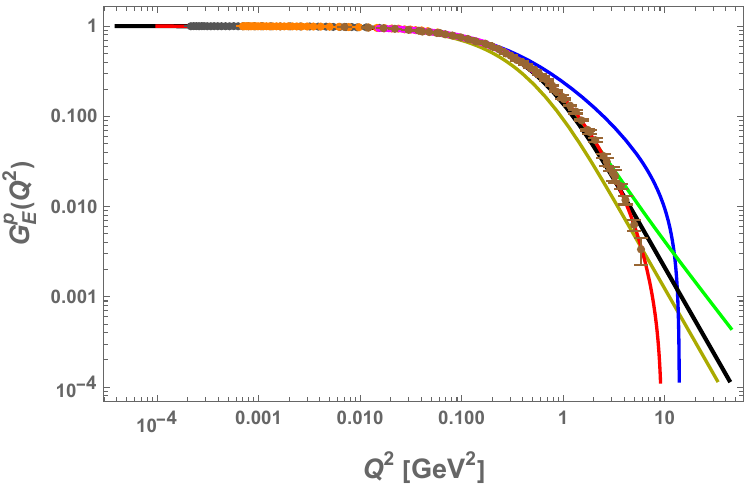}%
}\hfill
\subfloat[\label{fig_3pt6}]{%
  \includegraphics[height=5.5cm,width=.46\linewidth]{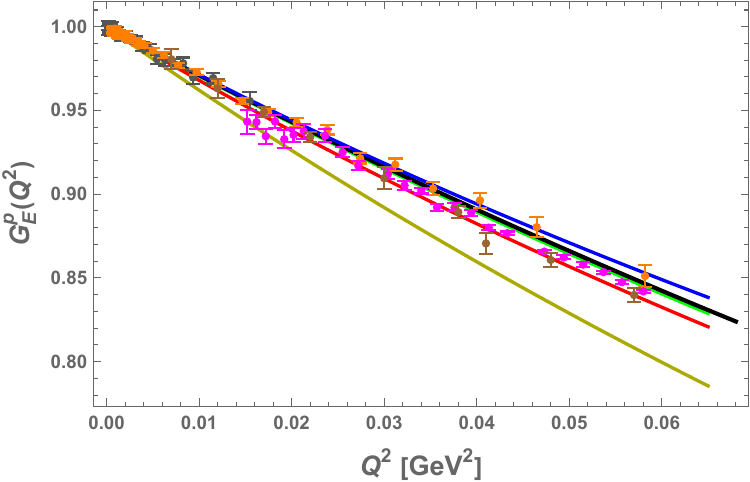}%
}\hfill
\subfloat[\label{fig_4pt1}]{%
  \includegraphics[height=5.5cm,width=.46\linewidth]{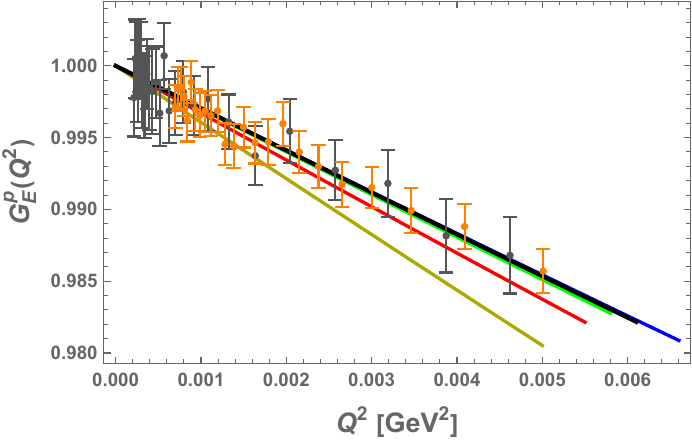}%
}
\caption{Sachs electric form factor of the proton at high and low momentum transfer. The black-solid curve is our soft-wall holographic QCD result with $\tau=3$ and $\kappa=0.402\,\text{GeV}$. The green-solid curve is our soft-wall holographic QCD result with $\tau={m_N^2}/{m_{\rho}^2}+1=2.465$ and $4\kappa^2=m_{\rho}^2=0.775^2\,\text{GeV}^2$. The  dark-yellow-solid curve is the soft-wall holographic QCD result for $\tau=3$ and $\kappa=0.350~\rm GeV$~\cite{Abidin:2009hr}. The solid-blue curve is the PRad fit to PRad data~\cite{Xiong:2019umf} (orange and gray data points). The solid-red curve is the Arrington fit to world data (without PRad and Mainz A1 data)~\cite{Arrington:2007ux} (brown data points). The magenta data points are the Mainz A1 data~\cite{Bernauer:2013tpr}. Note that in (c) our holographic prediction (solid-black curve) completely overlaps with the PRad fit to PRad data (solid-blue curve).}
\label{fig_GEP}
\end{figure*}

\section{Fixing the mass scale $\kappa$ and twist $\tau$, in the original-minimal soft-wall holographic QCD}\label{kappatau}
In this work, we first fix the twist parameter of the proton $\tau=3$,  to match the hard counting rule in the large $Q^2$ regime (see Appendix \ref{HARD} for more details). We then fix the mass scale $\kappa=0.402\,\text{GeV}$ by fitting the holographic Sachs magnetic form factor $G_{M}^{P}(Q)$~(\ref{magnetic}),  to the world experimental data (including the Mainz A1 data) on the Sachs magnetic form factor, see Table~\ref{table11}, and the solid-black curves in Fig.~\ref{fig_GMP}\,-\,\ref{fig_GEP}. 
 
In this work, we also fix $\kappa$, as was done in the original-minimal soft-wall holographic QCD \cite{Karch:2006pv}, by the mass of the $\rho$ meson using the relation $4\kappa^2=m_{\rho}^2$. Then, we fix $\tau$ by the mass of the nucleon using the relation $m^2_{N}=4\kappa^2(\tau-1)$. Therefore, for $m_{\rho}=0.775\,\text{GeV}$ and $m_N=0.938$, we find $\tau={m_N^2}/{m_{\rho}^2}+1=2.465$ and $\kappa=m_{\rho}/2=0.388\,\text{GeV}$, see Table~\ref{table11}, and the solid-green curves in Fig.~\ref{fig_GMP}\,-\,\ref{fig_GEP}.  

Our choice is in contrast to the earlier work of Abidin and Carlson \cite{Abidin:2009hr}, within the original-minimal soft-wall holographic QCD \cite{Karch:2006pv}, where they first fixed $\tau=3$ to match the hard counting rule in the large $Q^2$ regime, and then fixed $\kappa=0.350\,\text{GeV}$ (by simultaneously fitting to the experimental mass of the  rho meson and proton) which gives a smaller $m_{\rho}=2\kappa=0.700\,\text{GeV}$, and a larger mass of the nucleon $m_N=2\kappa\sqrt{\tau-1}=0.990\,\text{GeV}$, see Table~\ref{table11}, and the dark-yellow-solid curves in Figs.~\ref{fig_GMP}\,-\,\ref{fig_GEP}. 



In Fig.~\ref{fig_GMP}, we show the magnetic Sachs form factor of the proton. The black-solid curve is our soft-wall holographic QCD result with $\tau=3$ and $\kappa=0.402\,\text{GeV}$. The green-solid curve is our soft-wall holographic QCD result with $\tau={m_N^2}/{m_{\rho}^2}+1=2.465$ and $4\kappa^2=m_{\rho}^2=0.775^2\,\text{GeV}^2$. 
The  dark-yellow-solid curve is the soft-wall holographic QCD result for $\tau=3$ and $\kappa=0.350~\rm GeV$~\cite{Abidin:2009hr}. The solid-red curve is the Arrington fit to world data (without the Mainz A1 data)~\cite{Arrington:2007ux} (brown data points). The magenta data points are the Mainz A1 data~\cite{Bernauer:2013tpr}. Throughout, we have used the proton magnetic moment $\mu_p=G_{M}^{P}(0)=F_{1}^{P}(0)+F_{2}^{P}(0)=2.793$.

In Fig.~\ref{fig_GEPGMP}, we show the ratio of the Sachs electric and magnetic form factor of the proton at high and low momentum transfer. 
The black-solid curve is our soft-wall holographic QCD result with $\tau=3$ (fixed by the hard counting rule) and $\kappa=0.402\,\text{GeV}$ (extracted from the best fit to the world data (including Mainz A1 data) of Sachs magnetic form factor of proton). The green-solid curve is our soft-wall holographic QCD result with $\tau={m_N^2}/{m_{\rho}^2}+1=2.465$ and $4\kappa^2=m_{\rho}^2=0.775^2\,\text{GeV}^2$.  
The dark-yellow-solid curve is the soft-wall holographic QCD result for $\tau=3$ and $\kappa=0.350~\rm GeV$~\cite{Abidin:2009hr}. The solid-red curve is the Arrington fit \cite{Arrington:2007ux} to the world data~(brown data points) (note that the world data (brown data points) is without the Mainz A1 \cite{Bernauer:2013tpr} and the recent JLab recoil polarization experimental data \cite{Paolone:2010qc,Zhan:2011ji}). The magenta data points are the Mainz A1 data~\cite{Bernauer:2013tpr}.  The two cyan data points are from the recent high precision JLab recoil polarization experiment E03-104~\cite{Paolone:2010qc}. The dark-blue data points are from the recent high precision JLab recoil polarization experiment E08-007~\cite{Zhan:2011ji}.

In Fig.~\ref{fig_GEP}, we show the electric Sachs form factor of the proton. The black-solid curve is our soft-wall holographic QCD result with $\tau=3$ and $\kappa=0.402\,\text{GeV}$ (fixed using the world data of magnetic Sachs form factor of proton). The green-solid curve is our soft-wall holographic QCD result with $\tau={m_N^2}/{m_{\rho}^2}+1=2.465$ and $4\kappa^2=m_{\rho}^2=0.775^2\,\text{GeV}^2$. The dark-yellow-solid curve is the soft-wall holographic QCD result for $\tau=3$ and $\kappa=0.350~\rm GeV$~\cite{Abidin:2009hr}. The solid-blue curve is the PRad fit to PRad data~\cite{Xiong:2019umf} (orange and gray data points). The solid-red curve is the Arrington fit to world data (without PRad and Mainz A1 data)~\cite{Arrington:2007ux} (brown data points). The magenta data points are the Mainz A1 data~\cite{Bernauer:2013tpr}. Note that our holographic prediction (solid-black curve) completely overlaps with the PRad fit to the PRad data (solid-blue curve) in Fig.~\ref{fig_GEP}.c for $Q^2<0.006$.


\section{Electromagnetic radii}
\label{EMRadii}

\subsection{Proton}

The magnetic radius of the proton is defined as 
\bea\label{mrp11}
\left<r_M^2\right>_{p}=-6\bigg(\frac{d{\rm ln} G_M^{P}(Q)}{d Q^2}\bigg)_0 \,\hbar^2 c^2  \,,
\eea
with $\hbar c=0.197\,\text{GeV}\,\rm fm$. Using~(\ref{magnetic}) in~(\ref{mrp11}), we find   
\bea\label{mrp22}
&&\left<r_M^2\right>_{p}=(0.803~\rm fm)^2\nonumber\\
&&(\textbf{extracted from world data})\,,
\eea
with $\kappa$=0.402~\text{GeV} (fixed using the world data (including the Mainz A1 data) of the magnetic Sachs form factor of the
proton, see Fig.~\ref{fig_GMP}), and $\tau=3$ (fixed by the  hard scattering rule). Alternatively, we have
\bea\label{mrp33}
\left<r_M^2\right>_{p}=(0.791~\rm fm)^2\,,  
\eea
for $\kappa=m_{\rho}/2=0.388\,\text{GeV}$, and $\tau={m_N^2}/{m_{\rho}^2}+1=2.465$. Combining (\ref{mrp22}) and (\ref{mrp33}), we find our final magnetic radius of the proton
\bea\label{mrp44}
&&\left<r_M^2\right>_{p}=(0.803\pm 0.012~\rm fm)^2\nonumber\\
&&(\textbf{extracted from world data})\,,
\eea
in good agreement with the average of e-p scattering Mainz A1 data (with the average of the most flexible Spline and Polynomial fits)~\cite{Bernauer:2013tpr}, and world data (with Arrington and Sick (2015) world updated value~\cite{Arrington:2015ria} or equivalently with the extraction of Zhan, et al. (2011)~\cite{Zhan:2011ji}), respectively, which is $\left<r_M^2\right>_{p}^{\frac{1}{2}}=(0.777~\rm fm+0.867~\rm fm)/2=0.822~\rm fm$.

The charge radius of the proton is given by
\bea\label{crp11}
\left<r_C^2\right>_{p}=-6\bigg(\frac{d {\rm ln}G_E^{P}(Q)}{d Q^2}\bigg)_0\,\hbar^2 c^2  \,.
\eea
Using~(\ref{electric}) in~(\ref{crp11}), we find   
\bea\label{crp22}
\left<r_C^2\right>_{p}=(0.831~\rm fm)^2\,,  
\eea
for $\kappa$=0.402~\text{GeV}, and $\tau=3$, see Table~\ref{table11}\,-\,\ref{table22}. We also find
\bea\label{crp33}
\left<r_C^2\right>_{p}=(0.839~\rm fm)^2\,,  
\eea
for $\kappa$=0.388~\text{GeV}, and $\tau=2.465$, see Table~\ref{table11}. Combining (\ref{crp22}) and (\ref{crp33}), we find our final charge radius of the proton
\bea\label{crp44}
\left<r_C^2\right>_{p}=(0.831\pm 0.008~\rm fm)^2~(\textbf{prediction})\,,\nonumber\\  
\eea
see Fig.~\ref{fig_data11}.

\subsection{Neutron}

The magnetic radius of the neutron is defined as 
\bea\label{mrn11}
\left<r_M^2\right>_{n}=-6\bigg(\frac{d{\rm ln} G_M^{N}(Q)}{d Q^2}\bigg)_0 \,\hbar^2 c^2  \,.
\eea
Using~(\ref{magnetic}) in~(\ref{mrp11}), we find   
\bea\label{mrn22}
\left<r_M^2\right>_{n}=(0.817~\rm fm)^2\,,\nonumber\\ 
\eea
with $\kappa$=0.402~\text{GeV}, and $\tau=3$. We also find
\bea\label{mrn33}
\left<r_M^2\right>_{n}=(0.810~\rm fm)^2\,,  
\eea
for $\kappa=0.388\,\text{GeV}$, and $\tau=2.465$. Combining (\ref{mrn22}) and (\ref{mrn33}), we find our final magnetic radius of the  neutron
\bea\label{mrn44}
\left<r_M^2\right>_{n}=(0.817\pm 0.007~\rm fm)^2~(\textbf{prediction})\,,
\eea
in good agreement with the PDG world average~\cite{Zyla:2020zbs} $\left<r_M^2\right>_{n}^{\frac{1}{2}}=0.864~\rm fm$.

The charge radius of the neutron is given by
\bea\label{crn11}
\left<r_C^2\right>_{n}=-6\bigg(\frac{dG_E^{N}(Q)}{d Q^2}\bigg)_0\,\hbar^2 c^2  \,.
\eea
Using~(\ref{electric}) in~(\ref{crn11}), we find   
\bea\label{crn22}
\left<r_C^2\right>_{n}=(-0.094~\rm fm)^2\,,  
\eea
for $\kappa$=0.402~\text{GeV}, and $\tau=3$. We also find
\bea\label{crn33}
\left<r_C^2\right>_{n}=(-0.142~\rm fm)^2\,,  
\eea
for $\kappa$=0.388~\text{GeV}, and $\tau=2.465$. Combining (\ref{crn22}) and (\ref{crn33}), we find our final charge radius of the neutron
\bea\label{crn44}
\left<r_C^2\right>_{n}=(-0.094\pm 0.048~\rm fm)^2~(\textbf{prediction})\,.\nonumber\\  
\eea
in good agreement with the average of MAMI, JLab/Hall-A, and CLAS experiments (with H. Atac, et al. (2021) fit~\cite{Atac:2021wqj}), and the world data (with Kelly (2004) fit~\cite{Kelly:2004hm}), which is $\left<r_E^2\right>_{n}=(-0.110~\rm fm^2-0.112~\rm fm^2)/2=-0.111~\rm fm^2$. 

\begin{widetext}
We have summarized our result for the electromagnetic radii of the proton and neutron in Table~\ref{table22}.

\begin{table}[h]
\begin{center}
\begin{tabular}{||c||c|c||}
\hline\hline
Nucleon radii & Experimental values & Soft-wall AdS/QCD [\textbf{This work}] \\
\hline\hline
$\sqrt{\langle r^p_M\rangle^2}$ &  0.822 \,fm (e-p world average)&0.803 $\pm$ 0.012\, fm (\textbf{extracted})     \\
\hline
$\sqrt{\langle r^p_C\rangle^2}$ &  0.831\,fm  (PRad 2019)&0.831 $\pm$ 0.008\, fm (\textbf{predicted})    \\
\hline
$\sqrt{\langle r^p_C\rangle^2}$ &  0.841 \,fm ($\mu p$ Lamb shift)& 0.831 $\pm$ 0.008\, fm (\textbf{predicted})     \\
\hline
$\sqrt{\langle r^n_M\rangle^2}$ &  0.864 \,fm (PDG world average) &0.817 $\pm$ 0.007\, fm (\textbf{predicted}) \\
\hline
$\langle(r^n_C)^2\rangle$ &  -0.111 \,$\text{fm}^2$ (world average) & -0.094 $\pm$ 0.048 \,$\text{fm}^2$ (\textbf{predicted})     \\
\hline\hline
\end{tabular}
\end{center}
\caption{Comparison of the soft-wall holographic QCD electromagnetic radii of the proton and neutron with experiment. See text for the references of the quoted experimental values.}\label{table22}
\end{table}
\end{widetext}

\section{Conclusions}					
\label{Conclusion}
By fixing the twist parameter of the nucleon $\tau=3$, and $\kappa=0.402~\text{GeV}$, we have found a much improved description of the world data (including the Mainz A1~\cite{Bernauer:2010wm}, Jefferson Lab recoil polarization experiment E08-007~\cite{Zhan:2011ji}, and PRad data~\cite{Xiong:2019umf}), of both the electric and magnetic Sachs form factors (including their ratio) of the proton.
The agreement with the data range from low to high momentum transfers, all within the original-minimal soft-wall holographic QCD \cite{Karch:2006pv,Abidin:2009hr}, see Figs.~\ref{fig_GMP}\,-\,\ref{fig_GEP}. For example, from our electric Sachs form factor of the proton~(\ref{electric}), we have found the charge radius of the proton to be $\bf{r_{p}=0.831\,\pm\,0.008\,\,\text{fm}}$ in perfect agreement with the recent charge radius of the proton measured by the PRad collaboration at Jefferson Lab \cite{Xiong:2019umf}. Our theoretical error in the charge radius of the proton, stems from our alternative choice of $\tau={m_N^2}/{m_\rho^2}+1=2.465$, and $\kappa=m_\rho/2=0.388~\text{GeV}$, which gives the charge radius of proton to be $r_{p}=0.839\,\,\text{fm}$, see Table~\ref{table11}. 

To our knowledge, our holographic framework is the only theoretical work that produces a perfect match to the recent PRad collaboration measurement of the charge radius of the proton. We have also computed the charge and magnetic radii of the neutron, as well as the magnetic radius of the proton, with a good agreement to the experimental values, see Table~\ref{table22}. This is not surprising given the non-perturbative nature of the problem, and the phenomenological success of the original  soft-wall holographic QCD construction~\cite{Karch:2006pv,Abidin:2009hr,Grigoryan:2007my,Colangelo:2008us,Forkel:2007ru,Vega:2010ns,Braga:2011wa,Mamo:2021cle,Mamo:2019mka,Mamo:2021krl,Mamo:2021tzd}, with only three parameters: the mass scale $\kappa$, the twist parameter of the nucleon $\tau$, and the  $^\prime$t Hooft coupling constant $\lambda$ (also needed for high energy t-channel diffractive processes).    

Holography recovers the old dual description with all its phenomenological successes.
As we noted earlier, it is the most economical way of enforcing QCD symmetries, duality and 
crossing symmetries, all usually sought by dispersive analyses, and yet within a well defined 
and minimal organisational principle using field theory.
The present analyses and results are an illustration of that.

However,  holography provides
much more. Indeed, the bulk holographic description
in higher dimension allows for the assessment of any n-point function on the boundary,  using 
field theoretical methods through dual Witten diagrams, in the double limit of large $N_c$ and strong gauge coupling. 
It is the QCD string made user friendly.
It also provides for novel physics for processes at low parton-x~\cite{Mamo:2019mka},
and for scattering on dense systems as dual to black holes~\cite{Mamo:2018ync,Mamo:2019jia}.


\begin{acknowledgments}

K.M. is supported by the U.S.~Department of Energy, Office of Science, Office of Nuclear Physics, contract no.~DE-AC02-06CH11357, and an LDRD initiative at Argonne National Laboratory under Project~No.~2020-0020. I.Z. is supported by the Office of Science, U.S. Department of Energy under Contract No. DE-FG-88ER40388.

\end{acknowledgments}

\appendix

\begin{widetext}

\section{Soft-wall holographic QCD}
\label{Soft-wall holographic QCD}

A simple way to capture AdS/CFT duality in the non-conformal limit is  to model it using a slice of AdS$_5$ with various bulk fields with assigned 
anomalous dimensions and pertinent boundary values, in the so-called bottom-up approach which we will follow here using the conventions in our
recent  work in DIS scattering~\cite{Mamo:2019mka,Mamo:2021cle}.
 We consider AdS$_5$ with a soft wall with a background metric  $g_{MN}=(\eta_{\mu\nu},-1)R^2/z^2$ with the flat metric  $\eta_{\mu\nu}=(1,-1,-1,-1)$
 at the boundary.  Confinement will be described by  a harmonic background dilaton $\phi={\kappa}^2z^2$.

\subsection{Bulk vector mesons}

The vector mesons fields $L,R$  are  described by the bulk  effective action~\cite{Hirn:2005nr,Domokos:2009cq,Mamo:2021cle}

\begin{eqnarray}
\label{01}
S_{M}=&&-\frac 1{4g_5^2}\int d^5x e^{-\phi(z)}\sqrt{g}\,g^{MP}g^{NQ}{\rm Tr}\bigg({\cal F}^L_{MN}{\cal F}^L_{PQ}+{\cal F}^R_{MN}{\cal F}^R_{PQ}\bigg)
+\int d^5x\,\bigg(\omega^L_5({\cal A})-\omega^R_5({\cal A})\bigg)
\end{eqnarray}
with the Chern-Simons contribution

\be
\omega_5({\cal A})=
\frac{N_c}{24\pi^2}\int d^5x \,{\rm Tr}\left({\cal AF}^2+\frac 12 {\cal A}^3{\cal F}-\frac 1{10}{\cal A}^5\right)
\ee
Here ${\cal F}=d{\cal A}-i{\cal A}^2$ and ${\cal A}={\cal A}^aT^a$  with $T^0=\frac{3}{2}{\bf 1_2}$ and $T^i=\frac{1}{2}\tau^i$, with the form notation subsumed. 
Also the vector fields are given by $V=(R+L)/2$ and the  axial-vector  fields are given by $A=(R-L)/2$.
The coupling $g_5$ in (\ref{01})  is fixed by the brane embeddings in bulk, or phenomenologically as
${1}/{g_{5}^2}\equiv{N_c}/(12\pi^2)$~\cite{Cherman:2008eh}.

The flavor gauge fields solve

\begin{eqnarray}
\label{Solutionsgauge}
\Box  V^\mu  +z e^{\kappa^2 z^2}  \partial_z \Big( e^{-\kappa^2 z^2} \frac{1}{z}
\partial_z V^\mu \Big) \,=\, 0\qquad 
\Box V_z -  \partial_z \Big( \partial_\mu V^\mu \Big) \,=\, 0\,.
\end{eqnarray} 
subject to  the gauge condition

\begin{equation}
\label{gaugechoice}
 \partial_\mu V^\mu \,+\,
z e^{ \kappa^2 z^2} \partial_z \Big( e^{-\kappa^2 z^2} \frac{1}{z}
 V_z \Big) \,=\,0 \,,
\end{equation}
with the boundary condition $V_\mu (z, y) \vert_{z\to 0} \,=\, \epsilon_{\mu} (q)\, e^{-iq\cdot y}$.
The non-normalizable solutions are
\begin{eqnarray}
V_\mu (z, y) &=& \epsilon_\mu (q)\, e^{-iq\cdot y} \,\Gamma \bigg(1-\frac{q^2}{4\kappa^2} \bigg)\,\,\kappa^2z^2 
\,\,{\cal U} \bigg(1-\frac{q^2}{4 \kappa^2} ; 2 ; \kappa^2 z^2 \bigg)
\nonumber\\
V_z (z, y)  &=& \frac{i}{2} \,  \epsilon(q) \cdot q \,  e^{-iq\cdot y} \,\, \Gamma \bigg(1-\frac{q^2}{4\kappa^2 } \bigg)\,\, z \,\, 
{\cal U} \bigg(1-\frac{q^2}{4 \kappa^2} ; 1 ; \kappa^2 z^2 \bigg)\,,
\label{Gauge}
\end{eqnarray} 
with  $\,{\cal U} (a;b;w) \,$   the confluent hypergeometric functions of the second kind.  

The normalizable solutions (vector wavefunctions) are also given by~\cite{Grigoryan:2007my}

\be
\phi_n(z)=c_{n}\times\kappa^2z^2 L_n^1( \kappa^2z^2) \,,
\ee
with $c_{n}=\sqrt{{2}/{n+1}}$ which is determined from the normalization condition (for the soft-wall model with background dilaton $\phi=\kappa^2z^2$)
\be
\int dz\,\sqrt{g}e^{-\phi}\,(g^{xx})^2\,\phi_n(z)\phi_m(z)=\delta_{nm}\,,\nonumber\\
\ee
Here $L_n^{(\alpha)}(\xi)$ are the generalized Laguerre polynomials. Therefore, we have

\be
F_n=\frac 1{g_5}\bigg(-e^{-\phi}\frac{1}{z^\prime}\partial_{z^\prime}\phi_n(z^\prime)\bigg)_{z^\prime=\epsilon}=-\frac{2}{g_5}c_n(n+1)\kappa^2\,,\nonumber\\
\ee
with $\phi_n(z\rightarrow 0)\approx c_n\kappa^2z^2(n+1)$. Then, defining the decay constant as $$f_n=-{F_n}/{m_n}\,,$$  we have

\be
\phi_n(z)=\frac{f_n}{m_n}\times 2g_{5}\times\kappa^2z^2 L_n^1(\kappa^2z^2)\,,
\ee
as required by vector meson dominance (VMD).

We can also write the non-normalizable solution as sum over the normalizable solutions as~\cite{Grigoryan:2007my}

\be
V(q,z)= \sum_n \frac{-g_5F_n\phi_n(z)}{q^2-m_n^2}=g_5^2\times\sum_n \frac{f_n^2}{q^2-m_n^2}\times 2\xi L_{n}^1(\xi)\,, \label{vbbt2sw}
\ee
with $\xi=\kappa^2z^2$.

 \subsection{Bulk Dirac fermions}

The bulk Dirac fermion action in a sliced of AdS$_5$ is

\be
S_F=\frac 1{2g_5^2}\int d^{5} x \,e^{-\phi(z)}\,\sqrt{g}\,\Big(\mathcal{L}_{F1}+\mathcal{L}_{F2}\Big)+\frac 1{2g_5^2}\int d^4 x \sqrt{-g^{(4)}}\,\Big(\mathcal{L}_{UV1}+\mathcal{L}_{UV2}\Big)\,,\nonumber\\
\label{Action}
\ee
The Dirac and Pauli  contributions to ${\cal L}_{F1,2}$ are respectively

\bea
\label{fermionAction}
\mathcal{L}_{\rm Dirac1,2}&=&\bigg( \frac{i}{2} \overline{\Psi}_{1,2} e^N_A \Gamma^A\big(\overrightarrow{D}_N^{L,R}-\overleftarrow{D}_N^{L,R}\big)\Psi_{1,2}-(\pm M+V(z))\bar{\Psi}_{1,2}\Psi_{1,2}\bigg)\,,\nonumber\\
\mathcal{L}_{\rm Pauli1,2}&=&\pm 2g_5^2\times\eta\, \bar{\Psi}_{1,2} e^M_Ae^N_B\sigma^{AB}{\cal F}^{L,R}_{MN}\Psi_{1,2}\,,\nonumber\\
\eea
with  $V(z)={\kappa}^2z^2$,  $e^N_A=z \delta^N_A$, $\sigma^{AB}=\frac i2  [\Gamma^A,\Gamma^B]$, 
and  $\omega_{\mu z\nu}=-\omega_{\mu\nu z}=\frac{1}{z}\eta_{\mu\nu}$. The Dirac gamma matrices  
$\Gamma^A=(\gamma^\mu, -i\gamma^5)$ are chosen in the chiral representation.  They  satisfy the flat
anti-commutation relation $\{\Gamma^A,\Gamma^B\}=2\eta^{AB}$. 
The left and right covariant derivatives are defined as

\bea
\overrightarrow{D}_N^{X=L,R}=&&\overrightarrow{\partial}_N +\frac{1}{8}\omega_{NAB}[\Gamma^A,\Gamma^B]-iX_N^aT^a\equiv \overrightarrow{\mathcal{D}}_N-iX_N^aT^a \nonumber\\
\overleftarrow{D}_N^{X=L,R}=&&\overleftarrow{\partial}_N +\frac{1}{8}\omega_{NAB}[\Gamma^A,\Gamma^B]+iX_N^aT^a\equiv \overleftarrow{\mathcal{D}}_N+iX_N^aT^a 
\eea
The nucleon doublet  refers to 

\be
\Psi_{1,2}\equiv 
\begin{pmatrix} 
  \Psi_{p1,2}\\ 
  \Psi_{n1,2}
\end{pmatrix}\,.
\ee
The nucleon fields in bulk form an iso-doublet $p,n$ with $1,2$ referring to their  boundary chirality  $1,2=\pm=R,L$~\cite{Hong:2006ta}.  They are dual to the boundary sources
$\Psi_{p1,2}\leftrightarrow {\cal O}_{p,\pm}$  and $\Psi_{n1,2}\leftrightarrow {\cal O}_{n,\pm}$ with anomalous dimensions
$\pm M=\pm (\Delta-2)=\pm (\tau-3/2)$. 

The  equation of motions for the bulk Dirac chiral doublet  is

\bea
\label{EOM12}
&&\bigg(i e^N_A \Gamma^A D_N^{L,R} -\frac{i}{2}(\partial_N\phi)\, e^N_A \Gamma^A- (\pm M+V(z))\bigg)\Psi_{1,2}=0\,,
\eea
The normalizable solution to (\ref{EOM12}) are

\bea\label{SolutionFermions}
\Psi_1(p,z)&=&\psi_R(z)\Psi^0_{R}(p)+ \psi_L(z)\Psi^0_{L}(p)\nonumber\\
\Psi_2(p,z)&=&\psi_R(z)\Psi^0_{L}(p)+ \psi_L(z)\Psi^0_{R}(p)
\eea
with the normalized bulk wave functions

\be
&&\psi_R(z)=\tilde{n}_R {\xi^{\tau-\frac{3}{2}}}
			L^{(\tau-2)}_n(\xi)=\frac{\tilde{n}_R}{\kappa^{\tau-2}} z^{\frac{5}{2}}\xi^{\frac{\tau-2}{2}}L_n^{(\tau-2)}(\xi)\,,\nonumber\\
&&\psi_L(z)=\tilde{n}_L{\xi^{\tau-1}} 
			L^{(\tau-1)}_n(\xi)=\frac{\tilde{n}_L}{\kappa^{\tau-1}} z^{\frac{5}{2}}\xi^{\frac{\tau-1}{2}}L_n^{(\tau-1)}(\xi)\,,\nonumber\\
\ee
Here  $\xi=\kappa^2z^2$,  $L_n^{(\alpha)}(\xi)$, $\tilde{n}_R=\tilde{n}_L \kappa^{-1}\sqrt{\tau-1}$ are the generalized Laguerre, and $\tilde{n}_L=\kappa^{\tau}\sqrt{{2}/{\Gamma(\tau)}}$.  The free Weyl spinors $\Psi^0_{R/L}(p)= P_{\pm}u(p)$ and $\bar\Psi^0_{R/L}(p)=\bar u(p)P_{\mp}$,  and the free boundary spinors satisfy

\be
\bar u(p)u(p)=2m_N\qquad\qquad 
2m_N\times\bar u(p^{\prime})\gamma^{\mu}u(p)=\bar u(p^{\prime})\left[(p^{\prime}+p)^{\mu}+i\sigma^{\mu\nu}(p^{\prime}-p)_\nu \right] u(p)\,.
\ee
The fermionic spectrum Reggeizes  $m_n^2=4\kappa^2(n+\tau-1)$.
The assignments $1=+$ and $2=-$ at the boundary are commensurate with the substitutions
$\psi_{R,L}\leftrightarrow \mp \psi_{L,R}$ by parity.

Using the  Dirac 1-form currents

\begin{eqnarray}
J^{aN}_L=&&\frac{\partial\mathcal{L}_{\rm Dirac1}}{\partial L^a_N}=\overline{\Psi}_1 e^N_A \Gamma^A T^a\Psi_1\,,\nonumber\\
J^{aN}_R=&&\frac{\partial\mathcal{L}_{\rm Dirac2}}{\partial R^a_N}=\overline{\Psi}_2 e^N_A \Gamma^A T^a\Psi_2\,,
\end{eqnarray}
and Pauli 2-form currents

\begin{eqnarray}
J^{aMN}_L=&&\frac{\partial\mathcal{L}_{\rm Pauli1}}{\partial L^a_{MN}}=+2g_5^2\times\eta^{a}\overline{\Psi}_1 e^M_Ae^N_B \sigma^{AB}T^a\Psi_1\,,\nonumber\\
J^{aMN}_R=&&\frac{\partial\mathcal{L}_{\rm Pauli2}}{\partial R^a_{MN}}=-2g_5^2\times\eta^{a}\overline{\Psi}_2 e^M_Ae^N_B\sigma^{AB}T^a\Psi_2\,.
\end{eqnarray}
we can rewrite  (\ref{fermionAction}) with the explicit isoscalar ($a=0$) and isovector ($a=3$) contributions

\bea
\label{fermionAction2}
\mathcal{L}_{F1}+\mathcal{L}_{F2}\supset 
&&\frac{i}{2} \bar{\Psi}_1 e^N_A \Gamma^A\big(\overrightarrow{\mathcal{D}}_N-\overleftarrow{\mathcal{D}}_N\big)\Psi_1-(M+V(z))\bar{\Psi}_1\Psi_1 +\frac{i}{2} \bar{\Psi}_2 e^N_A \Gamma^A\big(\overrightarrow{\mathcal{D}}_N-\overleftarrow{\mathcal{D}}_N\big)\Psi_2-(-M+V(z))\bar{\Psi}_2\Psi_2\nonumber\\
&&+V_{N}^{0}J_{V}^{0N}+A_{N}^{0}J_{A}^{0N}+V_{N}^{3}J_{V}^{3N}+A_{N}^{3}J_{A}^{3N}+V^{0}_{MN}J_{V}^{0MN}+A^{0}_{MN}J_{A}^{0MN}+V^{3}_{MN}J_{V}^{3MN}+A^{3}_{MN}J_{A}^{3MN}\nonumber\\
\eea
with $J_{V,A}^{aN}=J_{L}^{aN}\pm J_{R}^{aN}$ and $J_{V,A}^{aMN}=J_{L}^{aMN}\mp J_{R}^{aMN}$.

\end{widetext}

\section{Electromagnetic form factors}
\label{Electromagnetic Form Factors}
The electromagnetic form factors for the nucleon  in the context of the soft wall model have been originally addressed in~\cite{Abidin:2009hr},
although with a different analysis at low momentum transfer than the one we will present and which  will fix  the numerical shortcomings 
they encountered.
Holographic QCD maps gauge invariant operators at the boundary with supergravity fields in bulk with pertinent anomalous dimensions.
The bulk srting theory is in general difficult to solve, but in the double limit of a large number of colors and strong gauge coupling 
$\lambda=g^2N_c$, it reduces to a weakly coupled supergravity in the classical limit in $AdS_5$ geometry, with a weak string coupling $g_{string}=g^2/4\pi$. 
The  n-point functions at the boundary of $AdS_5$ follow from variation of the
on-shell supergravity action in bulk with respect to the boundary values. The results are  tree-level Feynman graphs with fixed end-points on the
boundary also known as Witten diagrams~\cite{Nastase:2015wjb} (and references therein). 

\begin{widetext}
The electromagnetic form factor  for the nucleon can be extracted from the normalized boundary-to-bulk  Witten diagram for the
3-point function, with pertinent LSZ reduction
\begin{eqnarray}
{W}_{V}^{a\mu}(q)=\lim_{p_2^{2},p_1^2\to m_N^2}
{(p_2^{2}-m_{N}^ 2)(p_1^2-m_N^2)}\,\frac{\left<{\cal O}_{N}(-p_2)\,{J}_{V}^{a\mu}(q)\,{\cal O}_N(p_1)\right>}{F_N(p_2)F_N(p_1)}\,,
\end{eqnarray}
\end{widetext}
for the isoscalar $a=0$ and  the  isovector $a=3$ current $J_V^{a\mu}$,
The   nucleon source constant is  $F_N(p)=\left<0\left|{\cal O}_N(0)\right|N(p)\right>$.
The electromagnetic current form factors follow from the identification
\begin{widetext}
\bea
\label{DIRECTEM}
 \left<N(p_2)\left|{J}_{EM}^\mu(0)\right|N(p_1)\right>=
\bar u(p_2)\left(F_1(Q)\gamma^\mu + F_2(Q) \frac{i\sigma^{\mu\nu}q_\nu}{2m_n}\right)u(p_1)=\frac 13 {W}_{V}^{0\mu}(Q^2)+{W}_{V}^{3\mu}(Q^2)={W}_{EM}^\mu(Q^2)\nonumber\\
\eea
\end{widetext}
with $q^2=(p_2-p_1)^2=-Q^2$ (space-like). 
Here $N(p)$ refers to the U(2) proton-neutron doublet.

The vector currents ${J}_{V}^{a\mu}$ at the boundary 
are sourced by the dual  bulk vector fields  $V_{\mu}^{a}(Q,z\rightarrow 0)$. 
The Dirac and Pauli contributions to the direct part of the electromagnetic current can be extracted from the pertinent  bulk Dirac and Pauli parts of the action
in the soft wall model. The  Dirac part of the action in (\ref{fermionAction2})  yields
\begin{widetext}
\begin{eqnarray}
\label{InteractionActionEM}
S_{Dirac}^{EM} [i,f]=&& \frac{1}{2g_5^2}\, \int dz d^{4}y \sqrt{g} e^{-\phi} \,\frac{z}{R}
\nonumber\\
&&\times \bigg(\bar\Psi_{1f}  \gamma^N  \Big(\frac{1}{3}V_N^{0}T^0+ V_N^{3}T^3\Big)\Psi_{1i} 
+\bar{\Psi}_{2f} \, \gamma^N  \Big(\frac{1}{3}V_N^{0}T^0+ V_N^{3}T^3\Big)\Psi_{2i} \bigg) \,,\nonumber\\
=&&(2\pi )^4 \delta^4 ( p^{\prime} -p - q )\times F_N(p^{\prime})\times F_N(p)\nonumber\\
&&\times\,\frac{1}{2g_5^2}\times 2g_5^2\times e_{N}\times\bar{u}_{s_{f}}(p^{\prime}) \slashed\epsilon(q) u_{s_{i}}(p)\times\frac 12 \int \frac{dz}{z^{2M}}\, e^{-\phi}{\mathcal{V}}(Q,z)\, \left({\tilde{\psi}_L}^2(z)+{\tilde{\psi}_R}^2(z)\right)\,.\nonumber\\
\end{eqnarray}
\end{widetext}
Here $M=\tau-\frac{3}{2}$, $\phi=\kappa^2z^2$, and $\epsilon^\mu(q)$ is the polarization of the electromagnetic (EM) probe.
In the last equality in (\ref{InteractionActionEM}), we substituted the bulk gauge fields by (\ref{Gauge}), and the bulk fermionic currents in terms of the fermionic fields (\ref{SolutionFermions}).
The charge assignments are $e_{N}=1$ for the proton, and  $e_{N}=0$ for the neutron. We also use the nucleon (ground state) bulk wave functions given by~(\ref{SolutionFermions})
\be
\label{RXLX}
&&\tilde{\psi}_R(z)=n_R {\xi^{\tau-\frac{3}{2}}}L^{(\tau-2)}_0(\xi)\equiv n_R {\xi^{\tau-\frac{3}{2}}}\,,\nonumber\\
&&\tilde{\psi}_L(z)=n_L{\xi^{\tau-1}} 
L^{(\tau-1)}_0(\xi)\equiv n_L{\xi^{\tau-1}} 
\,,
\ee
with the generalized Laguerre polynomials $L_n^{(\alpha)}(\xi)$, $\xi=\kappa^2z^2$, $n_R = n_L \sqrt{\tau-1}$
and $n_L=\kappa^{-(\tau-2)}\sqrt{{2}/{\Gamma(\tau)}}$. We also use 
the non-normalizable solution to the bulk U(1) gauge field (the bulk-to-boundary vector propagator) (\ref{Gauge}) 
\begin{widetext}
\bea
\mathcal{V}(Q,z)
&=&g_5^2\times\sum_n \frac{-f_n^2}{Q^2+m_n^2}\times 2\xi L_{n}^1(\xi)=\xi \,\Gamma(1+a)\,\,{\cal U} (1+a; 2 ; \xi)=\xi\int_{0}^{1}\frac{dx}{(1-x)^2}x^{a}{\rm exp}\Big(-\frac{x\,\xi}{1-x}\Big)\,,\nonumber\\
\label{vps2sw}
\ee
\end{widetext}
with the decay constant of the vector meson resonances $f_n=\frac{\sqrt{2}\kappa}{g_5}$, linear Regge trajectory for mass of vector meson resonances $m_n^2=4\kappa^2(n+1)$, $a=Q^2/(4\kappa^2)$, and $\xi=\kappa^2z^2$. The bulk-to-boundary propagator satisfies ${\cal V}(0,z)={\cal V}(Q,0)=1$. 

The Dirac part of the electromagnetic current (\ref{DIRECTEM}) follows from  (\ref{InteractionActionEM}) by variation
\begin{widetext}
\bea
{W}_{EM(Dirac)}^\mu(Q^2)=\bar{u}_{s'}(p') \gamma^{\mu}u_{s}(p)\times e_N\times C_1(Q)\equiv \frac{1}{F_N(p')F_N(p)}\frac{\delta S_{Dirac}^{EM}}{\delta \epsilon_{\mu}(q)}
\eea
or more  explicitly
\bea
\label{C1Q}
C_1(Q) &=& \frac 12 \int \frac{dz}{z^{2M}}\, e^{-\phi}{\mathcal{V}}(Q,z)\, \left({\tilde{\psi}_L}^2(z)+{\tilde{\psi}_R}^2(z)\right)\label{C1} =g_5^2\times\sum_n \frac{-1}{Q^2+m_n^2}\times \left[f_n^2\int \frac{dz}{z^{2M}}\, e^{-\xi}\xi L_{n}^1(\xi)\, \left({\tilde{\psi}_L}^2(z)+{\tilde{\psi}_R}^2(z)\right)\right]\,,\nonumber\\
&=&g_5^2\times\sum_n \frac{-1}{Q^2+m_n^2}\times \mathcal{G}_{V_n\bar{B}B}=\int_{0}^1 dx\,\left(\tau\left(1-\frac{x}{2}\right)-\frac{1}{2}\right)\, x^{a}\,(1-x)^{\tau -2}=\frac{(a+2 \tau ) \Gamma (a+1) \Gamma (\tau )}{2 \Gamma (a+\tau +1)}\,,\nonumber\\
&=&\left(\frac{a}{2}+\tau \right)B(\tau ,a+1)\,,\nonumber\\
\eea
\end{widetext}
where, in the last line, we have defined the Euler beta function $B(x,y)=\frac{\Gamma(x)\Gamma(y)}{\Gamma(x+y)}$. In the second line, we have also defined the vector meson resonances $V_n$ coupling to the proton (baryon) $B$ as 
\be
\label{VBB}
\mathcal{G}_{V_n\bar{B}B}\equiv f_n^2\int \frac{dz}{z^{2M}}\, e^{-\xi}\xi L_{n}^1(\xi)\, \left({\tilde{\psi}_L}^2(z)+{\tilde{\psi}_R}^2(z)\right)\,,\nonumber\\
\ee
from which we can conclude that $\mathcal{G}_{V_n\bar{B}B}\sim N_c$, as expected in the large-$N_c$ limit.  Note that normalizing $C_1(0)=1$ for the proton, fixes $1+\mathcal{O}(N_c^{-2})=1$. The form factor readily crosses to time-like
(modulo $1/N_c$ widths), with all meson-baryon-antibaryon couplings (\ref{VBB}) fixed, in the double limit of large $N_c$ and strong
$^\prime$t Hooft coupling!  This  clearly shows the relationship  with constructions based on dispersive analyses~\cite{Lin:2021umk} (and references therein).


The nucleon electromagnetic form factors including the Pauli contribution follows a similar reasoning,
with the result for the proton 

\bea
\label{FP12}
F^{P}_1(Q)&=&C_1(Q)+\eta_{P}C_2(Q)\,,\nonumber\\
F^{P}_2(Q)&=&\eta_{P}C_3(Q)\,, 
\eea
and the neutron

\bea
\label{FN12}
F^{N}_1(Q)&=&\eta_{N}C_2(Q)\,,\nonumber\\
F^{N}_2(Q)&=&\eta_{N}C_3(Q)\,. 
\eea
The additional invariant form factors stem from the  Pauli contribution.
They are structurally similar to the Dirac one in~(\ref{C1Q}), with
\bea
\label{C23Q}
C_2(Q)&=&\frac 12 \int \frac{dz}{z^{2M}}\, e^{-\phi}\,z{\partial_z \mathcal{V}(Q,z)}\left({\tilde{\psi}_L}^2(z)-{\tilde{\psi}_R}^2(z)\right)\,,\nonumber\\
&=&\int_{0}^{1}dx\,\left(\tau x (\tau x+x-3)+1\right)\,x^{a}\,(1-x)^{\tau -2} \,,\nonumber\\
&=&\frac{a (a (\tau -1)-1) \Gamma (a+1) \Gamma (\tau )}{\Gamma (a+\tau +2)}\,,\nonumber\\
&=&(\tau -1)\frac{a (a (\tau -1)-1)}{(a+\tau ) (a+\tau +1)}B(\tau -1,a+1)\,,\nonumber\\
\eea
and
\bea
\label{C3Q}
C_3(Q)&=&2\int \frac{dz}{z^{2M}}\, e^{-\phi}\,m_Nz{\mathcal{V}(Q,z)}\tilde{\psi}_L(z)\tilde{\psi}_R(z)\,,\nonumber\\
&=&\int_{0}^{1}dx\,\left(4(\tau -1)\tau\right)\,x^{a}(1-x)^{\tau -1}\,,\nonumber\\
&=&\frac{4 (\tau -1) \tau  \Gamma (a+1) \Gamma (\tau)}{\Gamma (a+\tau +1)}\,,\nonumber\\
&=&4(\tau -1)\tau B(\tau ,a+1) \,,\nonumber\\
\eea
with $m_N^2=4\kappa^2(\tau-1)$.  Note that the contributions $C_{1,2}(Q)$ to the Dirac form factors
are chirality-spin preserving with $LL$ and $RR$ following the $\gamma^\mu$ assignment, while the contribution $C_3(Q)$ to the Pauli 
form factor is chirality-spin flipping with  $LR$ following the $\sigma^{\mu\nu}$ assignment.

\section{Hard scattering rule and the dual resonance model}
\label{HARD}

The hard scaling behavior for the electric form factor~\cite{Lepage:1980fj}  is recovered at high energy by fixing $\tau\rightarrow 3$.  
Indeed,   the  contribution $C_{1}(Q)$ to the Dirac form factor $F_1(Q)$ in
(\ref{C1Q})  with $a\gg 1$ and fixed $\tau$, asymptotes

\be
\label{HC1}
C_1(Q)=\left(\frac{a}{2}+\tau \right)\times B(\tau ,a+1)\approx \frac{1}{2}\frac{\Gamma(\tau)}{a^{\tau-1}}\,,
\ee
in agreement with the hard scattering rule for $\tau=3$. 
Similarly, the Pauli  contribution $C_{2}(Q)$ to the Dirac form factor $F_1(Q)$ in (\ref{C23Q}), at high momentum with $a\gg 1$ and fixed $\tau$, asymptotes   
\bea
\label{HC2}
C_2(Q)&=&(\tau -1)\frac{a (a (\tau -1)-1)}{(a+\tau ) (a+\tau +1)}B(\tau -1,a+1)\,,\nonumber\\
&\approx & (\tau-1)\frac{\Gamma(\tau)}{a^{\tau-1}}\,,
\eea
also in agreement with the hard scattering rule for $\tau=3$. Note that The Pauli term contribution $C_{3}(Q)$ to the Pauli form factor $F_2(Q)$ in
(\ref{C3Q}), at high momentum with $a\gg 1$ and fixed $\tau$, asymptotes to   
\be
\label{HC3}
C_3(Q)&=&4\tau(\tau -1) B(\tau ,a+1)\approx 4\tau(\tau -1)\frac{\Gamma(\tau)}{a^{\tau}}\,,\nonumber\\
\ee
which is subleading in $1/Q^2$.

This scaling law originates from the
contribution of the  R-spinor (\ref{RXLX}) in the form factor near the UV boundary, as it  dwarfs that of the L-spinor (\ref{RXLX}) 
when inserted  in the integral (\ref{C1Q}), 
with the bulk-to-boundary propagator  ${\cal V}(Q, z\sim  1/Q)\approx 1$. More specifically, the integrand in (\ref{C1Q}) gives
\be
\bigg(\frac{z}{z^{2M}}\bigg)\times \xi^{2\tau-3}(1+\xi)\approx \frac {z^{4\tau-6}}{z^{2\tau-4}}\approx \bigg(\frac 1{Q^2}\bigg)^{\tau-1}
\ee
Note that the Pauli contributions in (\ref{C23Q}) carry integrands with extra $z\sim 1/Q$ suppression  near the UV boundary,
and  are sub leading in $1/Q^2$. This is expected since the Pauli term is helicity-flipping.

Finally, we note the close relationship between  the nucleon form factor
obtained in the old resonance dual model~\cite{Frampton:1969ry}
\be
\label{OLD}
\frac{B(r_V, \frac 12 +a)}{B(r_V, \frac 12)}\\\nonumber
\ee
and the holographic form factors (\ref{HC1}-\ref{HC3}). In (\ref{OLD}),
the spin rho Regge trajectory is $$\frac 12+a=\frac 12+\alpha^\prime  t\,,$$  
with $r_V=\tau-1$ and  $\alpha^\prime=1/4\kappa^2$ the open string
or rho meson Regge slope.
Since the dual resonance description is rooted in the Veneziano amplitude for open string
scattering, it is not surprising that the holographic description which captures the dominant 
stringy excitations in bulk, yields similar results.

\bibliography{EMRADII}

\end{document}